# A New Semantic Web Approach for Constructing, Searching and Modifying Ontology Dynamically

Debajyoti Mukhopadhyay, Chandrima Chakrabarti, Sounak Chakravorty

Web Intelligence & Distributed Computing Research Lab
Green Tower, C-9/1, Golf Green
Calcutta 700095, India
{debajyoti.mukhopadhyay, hellochandrima, sounak.bhabook}@gmail.com

*Abstract-* ***Semantic web is the next generation web, which concerns the meaning of web documents* It has the immense power to pull out the most relevant information from the web pages, which is also meaningful to any user, using software agents.** *In today's world, agent communication is not possible if concerned ontology is changed a little. We have pointed out this very problem and developed an Ontology Purification System to help agent communication. In our system you can send queries and view the search results. If it can't meet the criteria then it finds out the mismatched elements. Modification is done within a second and you can see the difference. That's why we emphasis on the word dynamic. When Administrator is updating the system, at the same time that updation is visible to the user.*

*Index Terms-* ***Semantic Web, Search engine, Keyword based search, Ontology, Dynamic ontology, Resource Description Framework (RDF), RDF Schema(RDFS), Web Ontology Language (OWL).***

## 1. Introduction
For better understanding the meaning of data, we are using Ontology, which is the one of the major components of Semantic Web and Knowledge representation [1][2][3]. Ontology is the most intelligent way of describing a domain, which can be shared, visualized and understood easily. It gives us the freedom to search for any topic after specifying proper domain[4][5][6][7]. In our system after displaying the search results some verification is done. If any mismatching is found then the ontology is purified to give the end user the most relevant information. In the purification stage addition, modification or deletion are done as per the requirement. It seems to you that purification is very time consuming. But believe us our system is very fast and user-friendly. Your results may vary time to time due to our refreshing mechanism to represent you the most updated results. We think we can overcome the major drawbacks of the traditional search engines. Because their system based on the keyword based searching mechanism. These keyword based search engines most of the time fail to give the expected results, which are relevant to the query. Indexing quality is also very poor so you can't get important information unless you know the web address. In that case search engine is very much attentive to the spelling of the words not the meaning. Another thing is that from a huge pool of information it is most boring as well as time consuming task for a user to search his/her results. So naturally the needs of software agents are rising. When it is the situation then the development of semantic web is not far away. Where computer can understand the meaning of your query and return you the meaningful results. For better understanding we have developed specific ontology for a specific domain. We can modify our concepts and can make links to other domains, which have similar concepts.

## 2. Proposed Solution
In this section, the required procedure, algorithms and flowcharts are presented.

## 2.1 Procedure

```
Set input Character or String
Specify domain.
Match keywords within the specific
domain
If ontology not matched
Display "Mismatched ontology" message
 to user.
Apply Algorithm1 to find mismatched
elements.
Apply Algorithm3 to Purify Ontology.
Display results or crawl down links.
Else
Display results or crawl down links.
Stop.
```

## 2.2 Flow Chart

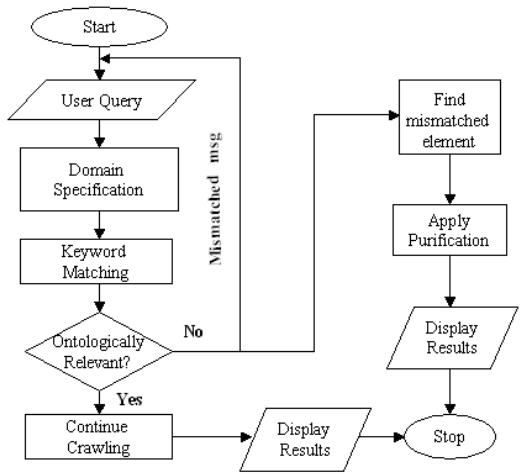

**Fig 1.** Search Engine Based on Dynamic Ontology

## 2.3 Our Ontology

There are so many ontologies available in the Internet. We have developed Theatre Ontology for the basis of our project. It's totally a new concept based on theatre domain.

The points which have influenced us to develop theatre ontology are specified below. They are as follows:
a. Vastness of theatre as an art form or subject.
b. User-friendly applications.
c. Proper organization of various aspects of
   theatre.
d. One of its kinds Ontology in the world.

   Combining all the above points we will be able to develop one of the finest definitions of cutting edge web site for the world. The pages to follow show the schematic diagram of the ontology for theatre.

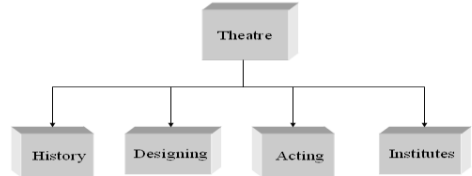

**Fig 2.** Basic Structure of Theatre Ontology

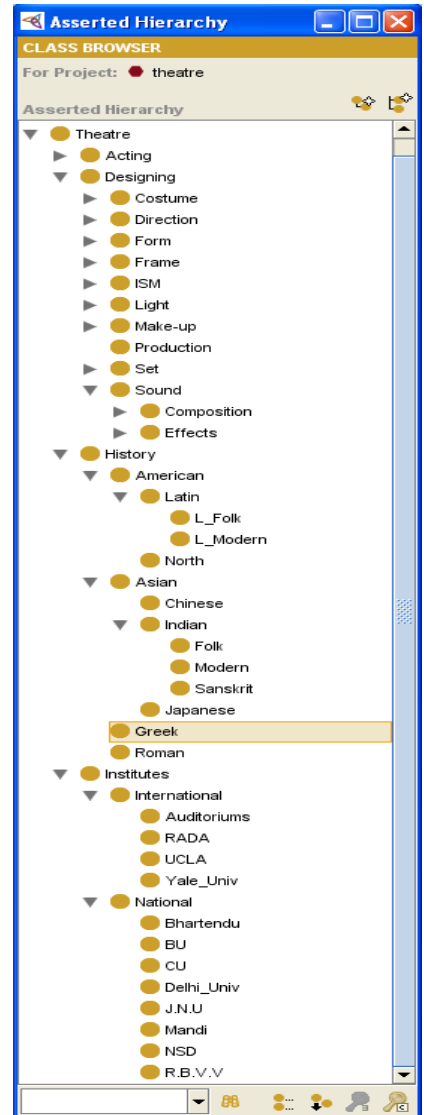

**Fig 3.** Here we are showing the part of our Theatre Ontology using **Prote`ge`**

## 2.4 Proposed Algorithms

At first we have to find out the mismatched nodes. We are assuming our ontology in a tree-like structure. From these tree we can find out the mismatched nodes.
Another new concept is we are calculating Mismatching Index.

If  Number of mismatched nodes =M

And  Total number of nodes=N

Then Mismatching index (mi) =**M/N**

For an ideal system there should not be any mismatching. That means for ideal system 'm' should be equal to zero. When we want to remove mismatching this should be kept in mind. Here black nodes are mismatched nodes. To find these nodes we have to traverse the tree. We have given each node a particular id. If ontological version remains same but still we found mismatching then by following id we can easily find out the mismatched nodes. Black nodes in Figure 4 indicate mismatched nodes. Our algorithm is as follow:

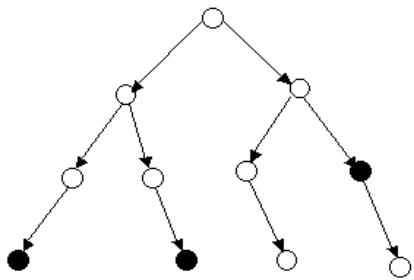

**Fig 4.**  Tree structure of Ontology

**Algorithm1 for finding number of mismatched nodes:**
```
Find   (Id,   Root,   Loc,   Par,   Left,
Right,Count)
Set Ptr=Root, Id=TRUE.
While (Ptr! = NULL)
{Compare version of two ontologies.
If (ver =FALSE)
Upgrade version and Exit.
Else
While (Id! =FALSE) Continue.
{Traverse the tree.]
If {Id [Node]<Id [Root]
Ptr=Left [Root]. Save=Root.}
Else {Ptr=Right [Root]. Save=Root.}
[Node found?]
For (Node=0;Node<n; Node++)
{Print the mismatched nodes.
Find number of mismatched nodes[M].
For (Count=0;Count<m;Count++){
Print total no. of mismatched nodes.
} Else
{[Search is unsuccessful]
 Set Ptr=NULL, Par=Save.}
}} Exit.
```

**Algorithm2 for counting total number of nodes:**
```
Count(Left,Right,Root,N)
Call Count(Left,Right,Root,N->Left)
Call Count(Left,Right,Root,N->Right)
Set N=N->Left+N->Right+1
Print Total number of Nodes [N].
Return.
```

**Algorithm3 for Purification of Ontology:**
```
Purify(Root,Id,mi,N,M,Left,Right,Par)
Set Ptr = Root, Id= TRUE.
Apply Algorithm1 to find the mismatched
nodes.
[If Node found]
Set Ptr=Node.
Calculate mismatching index (mi).
Count total number of nodes [N].
Apply Algorithm1 to find no of mismatched
nodes [M].
    mi=M/N;
While (mi! = 0) [while (mi) not equals to
zero.]
{Continue.
Compare Ids of two ontological nodes:
NodeO1 and NodeO2.
If {Id [NodeO1]! = Id [NodeO2]
Delete NodeO1.
Check immediate ancestor.
If
{Id[NodeO1]->Par=Id [NodeO2]->Par
Add new node.
If{ Id [NodeO1]<Id [NodeO1]->Par
Set Left[Par]=New
Else Set Right[Par]=New.
Set Id of new node.}
Print Updation Successful.}}
Else Updation Not Possible.
}
Exit.
```

## 2.5 Purification Techniques

In this phase of our proposed solution, after getting the mismatched elements, we need to purify the elements using addition, modification and deletion techniques. In the following three sub-sections, the steps are shown using three different flowcharts.

**A. Flowchart for adding new ontology element:**

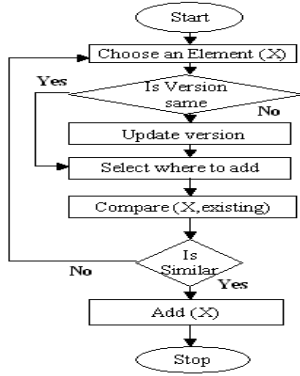

**Fig 5.** Adding new ontology element

**B. Flowchart for deleting existing ontology element:**

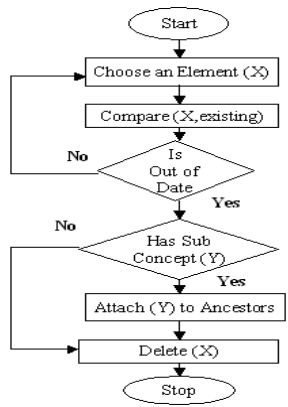

**Fig 6.** Deleting an ontology element

**C. Flowchart for modifying ontology element**:

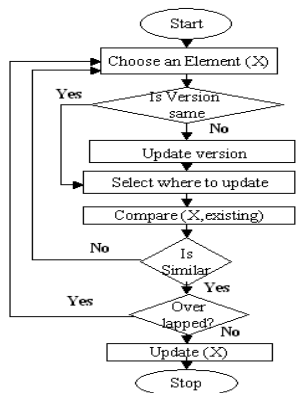

**Fig 7.** Modifying ontology element

## 3. Why OWL ?

OWL has a huge vocabulary to represent classes and relate them to their properties. Express relationship between classes (e.g. disjointWith, differentFrom, equivalentProperty, sameAs etc.) Express characteristics of properties (e.g. Object, Datatype, Annotation etc.), Fuctional, InverseFuctional, Symmetric, Transitive etc. and enumerated ,anonymous, deprecated classes. As OWL is the most powerful language than RDF or RDFS, we have built our ontology using OWL [8][9][10][11]. One must remember that the newly generated ontology is compatible with previous version or not. For that reason we should use **<owl:backwardCompatibleWith**> in Web Ontology Language (OWL). Also the old version of the ontology must be incompatible with the new ontology. In this case we should use **<owl:inCompatibleWith**>. We can also import some properties of the ontology in it's higher version. But if you are the owner of these two ontologies then and then only you can make this change[12].

## 4. Simulation

Now one can see our theatre ontology homepage in Figure 8. In the left side anyone can give query and view search results. Right side is for administrator who can update the ontology as per user requirement. That updates will be immediately visible to the user. After successful login an Administrator can insert any Ontology element easily as shown in Figure 10. If an Administrator wants to change any existing element he/she can do so using modify method (Figure 11). If any Ontology element is out of date or no longer used it is possible to delete the item by our system (Figure 12).

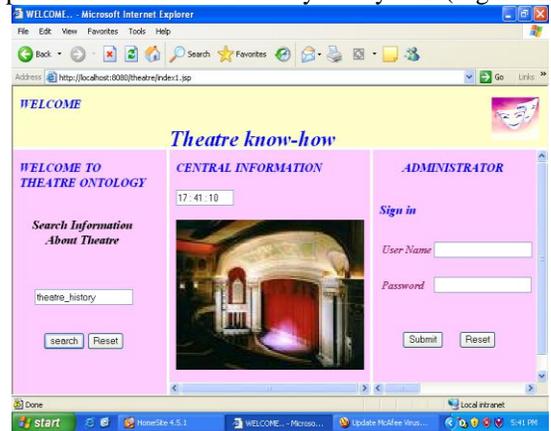

**Fig 8.** Homepage of Theatre Ontology based Search Engine.

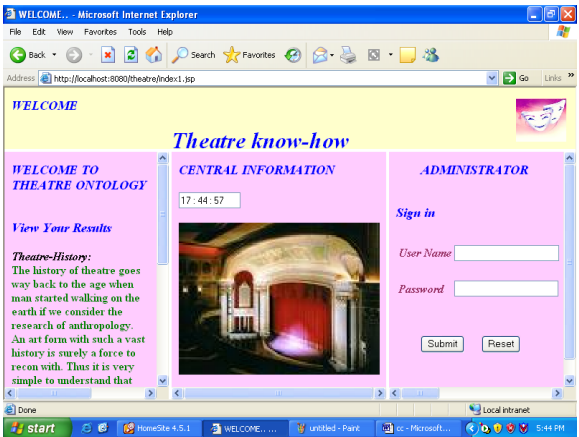

**Fig 9.** View search results

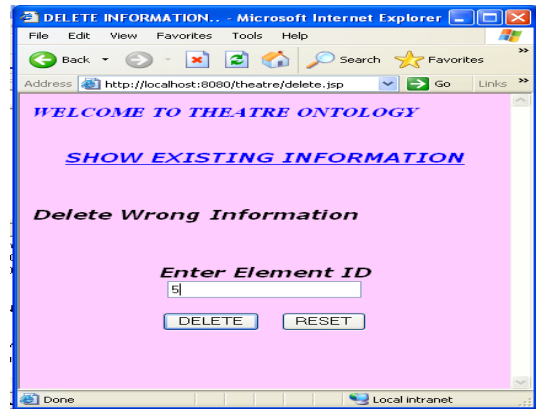

**Fig 12.** Delete mismatched element

Theatre ontology based search engine is much faster and accurate than any other search engine. We observe the performance of our Theatre ontology based search engine and traditional search engine for equal time span and the performance graphs are shown in Figure 13 and Figure 14 and find our approach with significant improvement.

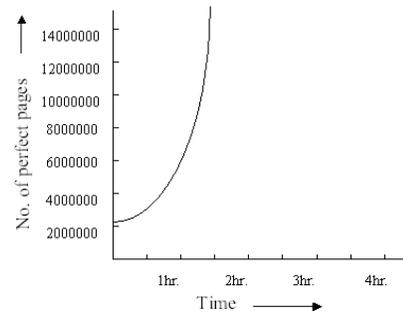

**Fig 13.** No. of perfect pages searched w.r.t Time with Ontology based Search Engine.

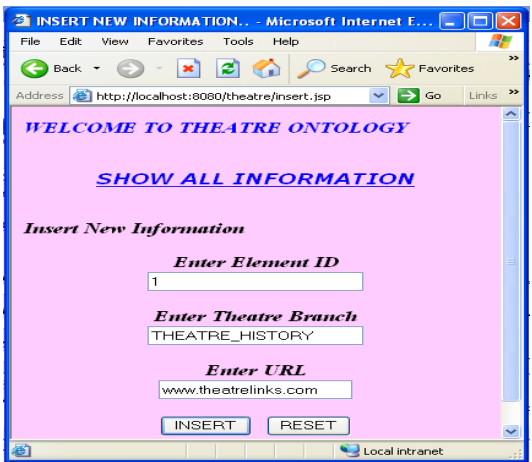

**Fig 10.** Insert new ontology element

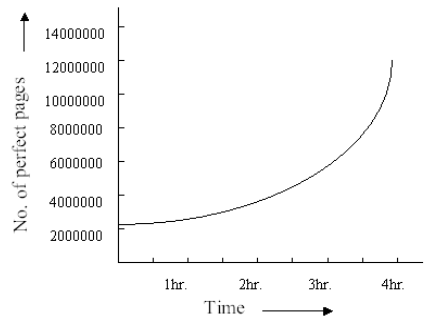

**Fig 14.** No. of perfect pages searched w.r.t Time with Traditional Search Engine.

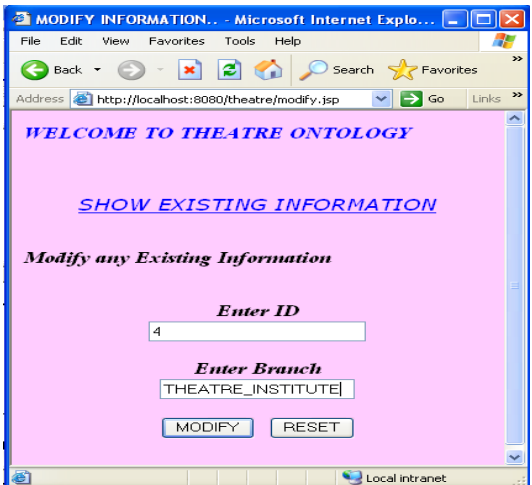

**Fig 11.** Modify existing element

## 5. Conclusions and Further Work

So we can find out the most challenging problem of today's agent communication i.e., Ontology mismatching. In this paper we have proposed algorithms for finding mismatched elements. At the same time apply algorithm to purify them. After the purification has been done successfully you can search through the ontology using our algorithm and can see the difference. We think Purification of Ontology will be the most appropriate method of dealing with this problem. It will help to make agent communication to make it successful and powerful too. There may be some limitations of our approach. However we rely on our approach that will enrich the development of Semantic Web.